\documentclass[11pt,leqno]{article}
\usepackage{geometry}
\usepackage{amsmath}
\usepackage{amsfonts}
\usepackage{amsthm}
\usepackage{indentfirst}

\numberwithin{equation}{section}

\theoremstyle{plain}
\newtheorem{theorem}{\indent\rm T\,h\,e\,o\,r\,e\,m\;}[section]

\theoremstyle{definition}
\newtheorem{definition}[theorem]{\indent\rm D\,e\,f\,i\,n\,i\,t\,i\,o\,n\;}

\theoremstyle{remark}

\makeatletter                                                  %
\renewcommand*{\@seccntformat}[1]{
	\csname the#1\endcsname\;-                                   %
}                                                              %
\renewcommand{\section}{\@startsection{section}{1}{0mm}        %
	{1.5\baselineskip}
	{1\baselineskip}
	{\indent\normalfont\normalsize\bfseries}
}                                                           %
\renewcommand*{\@seccntformat}[1]{
	\normalfont\bfseries\csname the#1\endcsname\;-               %
}                                                              %
\renewcommand\subsection{\@startsection                        %
	{subsection}{2}{0mm}
	{1.5\baselineskip}
	{1\baselineskip}
	{\indent\normalfont\normalsize\itshape}}
\renewcommand*{\@seccntformat}[1]{
	\normalfont\bfseries\csname the#1\endcsname\;-               %
}                                                              %
\renewcommand\subsubsection{\@startsection                     %
	{subsubsection}{2}{0mm}
	{1.5\baselineskip}
	{1\baselineskip}
	{\indent\normalfont\normalsize\texttt}}
\makeatother                                                   %
\begin{document}
	\thispagestyle{empty}


%
		
		\title{Monotone energy stability for Poiseuille flow in a porous medium}
		\author{Giuseppe Mulone\footnote{Universit\`{a} degli Studi di Catania (retired), Dipartimento di Matematica e Informatica, Viale Andrea Doria 6, 95125 Catania, Italy,  giuseppe.mulone@unict.it}}
		\date{}
		\maketitle
		
%
%
%
%
	
	\renewcommand{\thefootnote}{\fnsymbol{footnote}}
	
	\footnotetext{
		This research was partially supported by was partially supported by the following Grants: 2017YBKNCE of national project PRIN of Italian Ministry for University and Research,  PTR-DMI-53722122113 ``Analisi qualitativa per sistemi dinamici finito e infinito dimensionali con applicazioni a biomatematica, meccanica dei continui e termodinamica estesa classica e quantistica" of University of Catania, and the group GNFM of INdAM. }
	
	\renewcommand{\thefootnote}{\arabic{footnote}}
	\setcounter{footnote}{0}
	
	\vspace{0.6cm}
	\begin{center}
		\begin{minipage}[t]{11cm}
			\small{
				\noindent \textbf{Abstract.}
				We study the monotone energy stability of ``Poiseuille flow" in a plane-parallel channel with a saturated porous medium modeled by the Brinkman equation,  on the basis of an analogy with a 	magneto-hydrodynamic problem (Hartmann flow) (cf.  \cite{Hill.Straughan.2010}, \cite{Nield.2003}). 
				
				We prove that the least stabilizing perturbations, in the energy norm, are the two-dimensional spanwise perturbations. This result  implies a Squire theorem for  monotone nonlinear energy stability. Moreover, for Reynolds numbers less than the critical Reynolds number $R_E $ there can be no transient energy growth.
				\medskip
				
				\medskip
				
				\noindent \textbf{Mathematics~Subject~Classification:}
				76E05, 76S05.
				
			}
		\end{minipage}
	\end{center}
	
	\bigskip
	\textit{In memory of my dear friend Giampiero Spiga}
	\bigskip
	
	
\section{Introduction}
We study the problem of Poiseuille flow in a channel which is filled with a porous medium saturated with a linear viscous fluid. ``Due to the invention of man-made materials such as metallic foams which have a porosity close to one, and are of much use in the heat transfer industry, the porous – Poiseuille problem is of practical interest" (cf. \cite{Straughan.2008}). 

To study the flows in the porous medium, here we use the equations considered in \cite{Nield.2003} where an analogy is made with the Hartmann flows in MHD (cf. also \cite{Hill.Straughan.2010}, \cite{Straughan.2008},  and the references therein). 

Hill and Straughan, \cite{Hill.Straughan.2010},  study the linear instability and nonlinear stability of the base Poiseuille flow. They solve numerically the equivalent of the Orr-Sommerfeld eigenvalue
problem of fluid-dynamics. They also solve the nonlinear energy stability
eigenvalue problems for streamwise and spanwise disturbances. They use variational energy theory to derive a nonlinear global stability
threshold, $R_E$, and observe that ``the Euler-Lagrange equations
which arise are not in a form such that Squire’s theorem may be applied" (cf. \cite{Hill.Straughan.2010} p. 291). However, following \cite{Mulone.2023}, we shall prove here that this conclusion is not true and that the  critical Reynolds value for monotone energy stability is obtained on spanwise perturbations. The critical value they get on streamwise, as in Joseph \cite{Joseph.1976} and Busse \cite{Busse.1972}, underestimates the critical Reynolds number.

To prove that the critical Reynolds number for the nonlinear energy stability is obtained on the two-dimensional spanwise perturbations, after observing that a \textit{scale-invariant property},  \cite{Lamb.1924}, \cite{Lorentz.1907}, holds for the energy equation, we compare two maxima problems (see Section 3) in the space of kinematically admissible perturbations.  We prove that the two maxima coincide and the maximizing perturbations are two-dimensional spanwise perturbations. This implies a Squire's theorem for nonlinear energy stability.

The plan of the paper is as follows. 
In Section 2, we recall the basic equations given by \cite{Nield.2003} and \cite{Straughan.2008}, and write the base  Poiseuille flow. We give the definitions of streamwise and spanwise perturbations, and recall the instability results of Hill and Straughan \cite{Hill.Straughan.2010}.
In Section 3, we study monotone nonlinear energy stability and prove that the  Squire theorem holds for nonlinear monotone energy stability. In Section 4,  we  make  a conclusion.

\section{Basic equations and linear instability}

Consider a layer $\mathcal D_L= \mathbb R^2 \times [-L,L]$ filled with a saturated porous medium of Brinkman type. 
The basic equations are given in Nield [9], and in Straughan [11], Sect. 5.8.1. They are

\begin{eqnarray}\label{basicmotion-nd-a}
	\left\{ \begin{array}{l}
	\rho({\bf v}_t + 	{\bf v}\!\cdot\!\nabla{\bf v}) = - \nabla \bar p + \mu \Delta{\bf v} - \dfrac{\varphi \mu}{K} {\bf v}\\
		\nabla\!\cdot\!{\bf v}=0,
	\end{array}  \right.
\end{eqnarray}
where ${\bf v}={\bf v}(x,y,z,t) , \bar p= \bar p(x,y,z,t)$, and   $(x,y,z,t) \in \mathbb R^2 \times [-L,L]\times (0,+\infty)$.
In these equations $({\bf v}, \bar p)$ denote the velocity field and pressure,
 $\rho$  is the density, $\mu$ is the equivalent viscosity (for a Brinkman model), $\varphi$ is the porosity, and $K$ is the permeability (see \cite{Straughan.2008}) with $M_D= (\dfrac{\varphi}{Da})^{1/2}$, where $Da$ is the Darcy number defined by $Da= K/L^2$ ($M_D$ denotes the Darcy analogue of the Hartmann number in magnetohydrodynamics, see \cite{Nield.2003}).

As in \cite{Hill.Straughan.2010}, we use the scalings of $L, V$  and $L/V$ for length, velocity, and time. Equations \eqref{basicmotion-nd-a} may
be rewritten in non-dimensional form as
\begin{eqnarray}\label{basicmotion-nd-b}
	\left\{ \begin{array}{l}
	R({\bf v}_t + 	{\bf v}\!\cdot\!\nabla{\bf v}) = - \nabla \bar  p + \Delta{\bf v} - M^2 {\bf v}\\
		\nabla\!\cdot\!{\bf v}=0,
	\end{array}  \right.
\end{eqnarray}
where $R$ is the Reynolds number and $M^2$ is a non-dimensional (porous) quantity, given by

$$ R= \dfrac{\rho VL}{\mu}, \qquad   M^2= \dfrac{\varphi L^2}{K} .$$
The spatial domain is now $\mathcal D_1= \mathbb R^2 \times [-1,1]$, and the boundary conditions are 

$${\bf v}(x,y,\pm 1)=0 .$$

By choosing as $V$ the velocity at the center of layer $z=0$, we have that the base solution (or mean motion) is (see \cite{Takashima.1996}, (2.24))
\[ \label{BF} {\bf U}(x,y,z) = (U(z), 0,0), \qquad U(z) = \dfrac{\cosh M - \cosh (Mz)}{\cosh M -1}\]
which is the analogous of the Hartmann velocity in the MHD case, see \cite{Takashima.1996}.

As Straughan observed, \cite{Straughan.2008}, when the Darcy term disappears this should reduce to the classical Poiseuille solution for Navier-Stokes theory, namely $U(z) = (1-z^2)$, which one can recover from the previous formula in the limit $M  \to 0$.

The non-dimensional perturbation equations  are then

\begin{align}
	\label{pert-eqs}
	\left\{ \begin{array}{l}
		R(u_t  +  {\bf u}\!\cdot\!\nabla u+  U u_x+U' w)=  \Delta u - M^2  u- p_x\\[5pt]
	R(v_t +  {\bf u}\!\cdot\!\nabla v+  U v_x) = \Delta v - M^2  v- p_y\\[5pt]
	R(w_t +  {\bf u}\!\cdot\!\nabla w+  U w_x) = \Delta w - M^2  w - p_z\\	[5pt]
		u_x+v_y+w_z=0 ,\\	
	\end{array}  \right.
\end{align}
where $({\bf u}, p)$ are the perturbations to the velocity and the pressure fields, ${\bf u}=(u,v,w)$, with the boundary conditions 
$${\bf u}(x,y,\pm 1, t)=0 .$$

\begin{definition} {\rm We define \emph{streamwise} (or longitudinal) perturbations the perturbations 	${\bf u}, p$ which do not depend on $x$.} 
\end{definition}

\begin{definition} {\rm We define \emph{spanwise} (or transverse) perturbations the perturbations 	${\bf u}, p$ which do not depend on $y$. }  The two-dimensional spanwise perturbations are the spanwise perturbations with $v=0$.
\end{definition}

The linear instability of the base motion has been studied in \cite{Nield.2003} and a table of critical Reynolds values is given there. In  \cite{Hill.Straughan.2010} the linear instability has been studied by using the Squire theorem \cite{Squire.1933} and by solving the analogue of the Orr-Sommerfeld equation for the Brinkman porous theory.
In \cite{Hill.Straughan.2010} the authors showed plots of neutral curves for $M$ values from $0$ to $10$ (\cite{Hill.Straughan.2010}, Fig. 1 and Table 1): they  get critical linear Reynolds numbers ranging from $5772$ to $440223$. Due to Squire's theorem, these critical values are obtained on \textit{spanwise perturbations}.

\section{Nonlinear monotone energy stability}

Here we  study \textit{ the  nonlinear monotone  energy stability} with the Lyapunov second method.

Assume that both ${\bf u}$ and $\nabla p $ are $x,y$-periodic with periods $2\pi/a$ and $2\pi/b$ in the $x$ and $y$ directions, respectively,  with wave numbers $( a, b) \in \mathbb R^2_+$  . In the following it suffices therefore to consider functions over the periodicity cell 
$$\Omega= [0, \frac{2\pi}{a}]\times [0, \frac{2\pi}{ b}] \times [- 1, 1] .$$
As the basic function space, we take $L_2(\Omega)$, which is the space of square-summable functions in $\Omega$ with the scalar product denoted by
$$(g,h) = \int_0^{\frac{2\pi}{a}} \int_0^{\frac{2\pi}{ b}} \int_{-1}^1 g(x,y,z) h(x,y,z) dxdydz, $$ 
and the corresponding  norm $\Vert g \Vert = (g,g)^{1/2}.$

Taking into account the solenoidality of ${\bf u}$ and the boundary condition, we  write the Reynolds-Orr energy equation (sometimes it is called energy identity), \cite{Reynolds.1895}, Hill and Straughan \cite{Hill.Straughan.2010}

\begin{align}
	\label{Energy}
	R \dot E= -R (U'w,u) -  [\Vert \nabla u \Vert^2+\Vert \nabla v \Vert^2+\Vert \nabla w \Vert^2]- M^2  [\Vert u \Vert^2+\Vert  v \Vert^2+\Vert  w \Vert^2] ,
\end{align}
where the energy $E$ is defined by
\begin{equation} \label{EN-EQ}
E(t) = \dfrac{1}{2}[\Vert u \Vert^2 +  \Vert v \Vert^2 + \Vert w \Vert^2 ]. 
\end{equation}

Let us recall some definitions of nonlinear energy stability
(see \cite{Joseph.1976}, \cite{Schmid.Henningson.2001}).

\begin{definition} A base motion $U(z)$ is \emph{monotonically stable} in the energy norm $E$ of a disturbance, and $R_E$ is the critical Reynolds number, if the time orbital derivative  $\dot E$ of the energy is always less than zero
when $ R < R_E $. In particular the stability is \emph{ monotonic (or monotone) and
exponential} if there is a positive number $\alpha$ such that 
$$E(t) \le E(0) e^{-\alpha t}$$ for any $t\ge 0$ and $R<R_E$ .
\end{definition}

\begin{definition} A base motion $U(z)$ to the Navier–Stokes equations is \emph{globally stable} to perturbations if the perturbation energy $E$ satisfies
$$\lim_{t\to +\infty} E(t)=0 .$$
\end{definition}

We first note that, in the fluid-dynamics case, Lorentz \cite{Lorentz.1907}  made this observation (see Lamb \cite{Lamb.1924}, p. 640) : ``One or two consequences of the energy equation may be noted. In the first place, the relative magnitude of the two terms on the right-hand side  is unaffected if we reverse the signs of $u, v, w$, or if we multiply them by any constant factor. The stability of a given state of mean motion should not therefore depend on the \textit{scale} of the disturbance".
The same \textit{scale invariance} property holds  in the energy equation in the case under examination for Poiseuille flow in a porous medium.

This \textit{scale invariance} property in particular means that the term  $-(U'u,w)$ has a positive or negative sign and cannot change by exchanging the sign of only one of the two components $u$ and $w$. 

For perturbations with  $ -(U'w,u) \le 0$ and ${\bf u} \not= 0$, then $\dot E < 0$.

\noindent  If $ -(U'w,u) > 0$, from \eqref{Energy}, we easily have

\begin{eqnarray}
	\begin{array}{l}\label{ineq-}
		R \dot E= R \left(\dfrac{-(U'w,u)}{\Vert \nabla {\bf u} \Vert^2+ M^2 \Vert {\bf u}\Vert^2} - \dfrac{1}{{R}}\right) [\Vert \nabla {\bf u} \Vert^2 + M^2\Vert {\bf u} \Vert^2  ] \le \\[3mm]
		\le R \left(m - \dfrac{1}{R}\right)[\Vert \nabla {\bf u} \Vert^2 + M^2\Vert {\bf u} \Vert^2  ],
	\end{array}
\end{eqnarray}
where
\begin{eqnarray} \label{max1}
	\dfrac{1}{R_E} = m= \max_{\cal S} \dfrac{-(U'w,u)}{\Vert \nabla {\bf u} \Vert^2+ M^2 \Vert {\bf u}\Vert^2}, 
\end{eqnarray}
and $\cal S$ is the space of the   {\textit{kinematically admissible fields}}  
\begin{eqnarray} \nonumber
	\begin{array}{l}
		\label{spaceS}
		{\cal S}= \{u, v, w \in H^1(\Omega), \; u=v=w=0 \quad \hbox{on the boundaries,}\\[3mm] \hbox{ periodic in \textit{x}, and \textit{y},}  \quad u_x+v_y+w_z=0,\quad  {\bf u} \not= 0\},
	\end{array}
\end{eqnarray}
satisfying the \textit{scale invariance property}, 
with $H^1(\Omega)=W^{1,2}(\Omega)$ the usual Sobolev space: the subset of functions ${\displaystyle h} \in {\displaystyle L_{2}(\Omega )}$ such that ${\displaystyle h}$ and its weak derivatives up to order ${\displaystyle 1}$ have a finite $L_2$-norm.

As in \cite{Mulone.2023}, it can be proved, and easily verified, that when $ -(U'w,u) > 0$ for any ${\bf u} \in {\cal S}$ we have

$$\dfrac{-(U'w,u)}{\Vert \nabla {\bf u} \Vert^2+ M^2 \Vert {\bf u}\Vert^2} \le
\dfrac{-(U'w,u)}{\Vert \nabla {u} \Vert^2+ \Vert \nabla {w} \Vert^2+ M^2 (\Vert { u}\Vert^2+\Vert { w}\Vert^2 ) } $$
and that the maximum of the second ratio is obtained for $v=0$, $\dfrac{\partial}{\partial y}=0$, and that it coincides with the maximum of the first ratio.

The Euler-Lagrange equations of the maximum

\begin{equation} \label{max2} m= \max_{\cal S} \dfrac{-(U'w,u)}{\Vert \nabla {u} \Vert^2+ \Vert \nabla {w} \Vert^2+ M^2 (\Vert { u}\Vert^2+\Vert { w}\Vert^2 ) }\end{equation}
are given by the system

\begin{align}
	\label{elwz-comp}
	\begin{cases}
		-U'{w}+2 m (\Delta {u} - M^2 u)&= \lambda_x\\
	\quad	0&= \lambda_y\\
		-U'{u}  + 2 m (\Delta {w}- M^2 w) &= \lambda_z ,
	\end{cases}
\end{align}
where $\lambda$ is a Lagrange multiplier. Searching solutions $$(u,v,w,\lambda)= \bar H(z) e^{i(ax+by)},$$ with $a$ and $b$ wave numbers and $\bar H(z)= \bar u(z), \bar v(z), \bar w(z), \lambda(z)$ respectively, we have: $b=0$. Therefore, $u,v,w, \lambda$ are functions of $x$ and $z$, and $v_y=0$, $u_x+w_z=0$. Since the space of fields $(u(x,z),v(x,z),w(x,z))$ in $\cal S$, with $u_x+w_z=0$,  is a subspace of $\cal S$, the two maxima: \eqref{max1} and the maximum on the first side of equation \eqref{max2} are equal,  and $v_x=v_z=0$. These equations together with $v_y=0$ and the boundary conditions imply $v=0$.
We observe that: 

a) the maximizing field $(\bar u(x,z), 0, \bar w(x,z))$ satisfies the invariance scale property, 

b) if $\lambda=0$, then by solving   system \eqref{elwz-comp} with boundary conditions $w=w_z=w_{zz}=0$ on $z=\pm 1$,  we find that $u=w=0$ in $\Omega$; in this case, from the energy equation immediately it follows that the perturbations $(0, v(x,z), 0)$ are always stabilizing.

The critical Reynolds number ${R}_E$ is given by the (generalized) Orr's equation, see \cite{Hill.Straughan.2010}, formula (7), 
\begin{align} 
	\label{Orr-eq}
	{R}_E (U'' w_x+2U' w_{xz})+ 2(\Delta \Delta w-M^2 \Delta w)=0,
\end{align}	with b.c. $w=w'=0$.

This equation has been solved in \cite{Hill.Straughan.2010}, where a graph of $R_E$ against $M$ has been reported (in the caption of  Fig. 3 in \cite{Hill.Straughan.2010} the wave number $a$ has been written incorrectly instead of $M$).

From inequality \eqref{ineq-} and the Poincaré's inequality, we obtain
\begin{align}
	\label{time-en11}
	\begin{array}{l}
		\dot E	\le  \left( \dfrac{1}{{R}_E} - \dfrac{1}{{ R}}\right)\left[\dfrac{\pi^2}{2} + 2M^2\right] E .
	\end{array}
\end{align} 

By integrating this inequality, we get:
\begin{theorem}
Assuming ${R} <{{R}_E}$, the  basic ``Poiseuille flow" ${\bf U}$ is nonlinear monotone exponentially stable according to the classical energy:
\[ E(t) \le E(0) \exp\left\{{\left( \dfrac{1}{{R}_E} - \dfrac{1}{{ R}}\right)}(\dfrac{\pi^2}{2}+2M^2) t\right\}, \quad \forall t \ge 0.
\]
\end{theorem}

A consequence of this result is that a \textit{Squire theorem}, \cite{Squire.1933}, holds for nonlinear energy stability of Poiseuille flow in a porous medium: the less stabilizing perturbations in the energy norm are the two-dimensional spanwise  perturbations. 

\section{Conclusion}

We have studied the monotone energy stability of ``Poiseuille flow" in a plane-parallel channel with a saturated porous medium modeled by the Brinkman equation,  on the basis of an analogy with a magneto-hydrodynamic problem (Hartmann flow),\cite{Nield.2003}, (see also \cite{Mulone.2023a}).

We have proved that the least stabilizing perturbations, in the energy norm, are the two-dimensional spanwise perturbations. This result  implies a Squire theorem for  monotone nonlinear energy stability. Moreover, for Reynolds numbers less than the critical Reynolds number $R_E $ there can be no transient energy growth. 

This result appears to be in contrast with the critical value found by Hill and Straughan \cite{Hill.Straughan.2010} on the streamwise perturbations. But we note that, as in Joseph \cite{Joseph.1976} and Busse \cite{Busse.1972}, the critical value obtained by Hill and Straughan \cite{Hill.Straughan.2010} on the streamwise perturbations underestimates the nonlinear critical Reynolds number because it is calculated when in the energy equation the term $-(U'w,u)$ is replaced by its absolute value $\vert -(U'w,u) \vert$ on the streamwise. Finally we observe that streamwise perturbations do not verify the \textit{scale invariance property} unless $u=0$.
	
\vspace{0.5cm} \indent {\it
A\,c\,k\,n\,o\,w\,l\,e\,d\,g\,m\,e\,n\,t\,s.\;} The author
acknowledge support by the following Grants: 2017YBKNCE of national project PRIN of Italian Ministry for University and Research, PTR DMI-53722122146 "ASDeA" of the University of Catania. The author also thank the group GNFM of INdAM for financial support.

\bigskip
\begin{center}

\end{center}

	\bigskip
	\bigskip
	\begin{minipage}[t]{10cm}
		\begin{flushleft}
			\small{
				\textsc{Giuseppe Mulone}
				\\*University of Catania (retired),
				\\*Department of Mathematics and Computer Science
				\\*Viale andrea Doria, 6
				\\* Catania, 95125, Italy
				\\*e-mail: giuseppe.mulone@unict.it
				}
		\end{flushleft}
	\end{minipage}
	
\end{document}